\documentclass[aps,prb,superscriptaddress,preprintnumbers,
showpacs,legalpaper,twoside,twocolumn]{revtex4}
\usepackage{bm}
\usepackage{graphicx}
\usepackage{amsfonts}
\usepackage{amsmath}
\usepackage{amssymb}

\begin{document}

\title{Quantum disorder and Griffiths singularities in
bond-diluted two-dimensional Heisenberg
antiferromagnets}

\author{Rong Yu}
\affiliation{Department of Physics and Astronomy, University of
Southern California, Los Angeles, CA 90089-0484}
%\affiliation{Kavli Institute for Theoretical Physics, University of California,
%Santa Barbara, CA 93106-4030}
\author{Tommaso Roscilde \footnote{Current address: Max-Planck-Institut
f\"ur Quantenoptik, Hans-Kopfermann-str. 1, D-85748  Germany.}}
\affiliation{Department of Physics and Astronomy, University of
Southern California, Los Angeles, CA 90089-0484}
\author{Stephan Haas}
\affiliation{Department of Physics and Astronomy, University of
Southern California, Los Angeles, CA 90089-0484}

% 75.10.Jm - Quantized spin models
% 64.60.Ak - Renormalization-group, fractal, and percolation studies of
%            phase transitions
% 75.10.Nr - Spin-glass and other random models
% 75.40.-s - Critical-point effects, specific heats, short range order
% 75.40.Cx - Static properties
% 75.40.Mg - Numerical simulation studies

\begin{abstract}
We investigate quantum phase transitions in the spin-$1/2$
Heisenberg antiferromagnet on square lattices with
\emph{inhomogeneous} bond dilution. It is shown that quantum
fluctuations can be continuously tuned by inhomogeneous bond
dilution, eventually leading to the destruction of long-range
magnetic order on the percolating cluster. Two multicritical points
are identified at which the magnetic transition separates from the
percolation transition, introducing a novel quantum phase
transition. Beyond these multicritical points a quantum-disordered
phase appears, characterized by an infinite percolating cluster with
short ranged antiferromagnetic order. In this phase, the
low-temperature uniform susceptibility diverges algebraically with
non-universal exponents. This is a signature that the novel
quantum-disordered phase is a \emph{quantum Griffiths phase}, as
also directly confirmed by the statistical distribution of local
gaps. This study thus presents evidence of a genuine quantum
Griffiths phenomenon in a two-dimensional Heisenberg
antiferromagnet.
\end{abstract}

\maketitle

\section{\label{intro}Introduction}

Quantum phase transitions in low-dimensional
quantum Heisenberg
antiferromagnets (QHAFs) have been the subject of
extensive investigations during the last two decades
\cite{Chakravartyetal88, Sachdev99}. The transition
from a \emph{renormalized classical} to a
\emph{quantum-disordered} state can be triggered
by various parameters, including lattice dimerization,
frustration and applied field. More recently, special
interest has focused on phase transitions
driven by geometric randomness of the lattice \cite{Vajketal02, Sandvik02A},
including site and bond disorder. Strong geometric
disorder not only breaks translational invariance and perturbs
the ground state of the pure system, but it can also destabilize
renormalized classical phases with long-range order (LRO)
and drive the system to novel disordered phases.

Various one-dimensional QHAFs with bond disorder have been found to
display unconventional quantum phases\cite{Maetal79,Fisher94}. For
example, the undimerized QHAF chain is driven into a
\emph{random-singlet phase} \cite{Fisher94} with algebraically
decaying spin-spin correlations.
%v2!
%and power-law logarithmic divergent
%susceptibility for low temperatures.
The low temperature susceptibility diverges as $1/[Tlog^2T]$ in this
phase, independent of the details of the bond disorder. In contrast,
a dimerized chain shows a \emph{quantum Griffiths
phase}\cite{Hymanetal96} beyond a critical disorder strength, with
exponentially decaying spin-spin correlations and a non-universal
power-law divergent susceptibility. In two dimensions, the clean
QHAF develops antiferromagnetic (N\'eel) LRO at zero temperature
\cite{Manousakis91}. This introduces the intriguing possibility of a
genuine order-disorder transition, driven by lattice randomness,
i.e. from the N\'eel phase into one of the above unconventional
disordered phases \cite{note}.

Site and bond dilution of the square lattice QHAF have been the
focus of several recent studies \cite{Vajketal02, Sandvik02A,
Chernyshevetal02, Muccioloetal04, Bray-AliM04}, motivated by
experiments on antiferromagnetic cuprates doped with nonmagnetic
impurities \cite{Vajketal02, Cortietal95, Carrettaetal97}. From a
geometric point of view, bond and site dilution reduce the
connectivity of the lattice, ultimately leading to a percolative
phase transition \cite{StaufferA94} beyond which the system is
broken up into finite clusters. In a classical spin system, this
percolation transition is coupled to a magnetic transition with the
same critical exponents since spontaneous magnetic order cannot
survive beyond the percolation threshold. In a quantum spin system,
on the other hand, a progressive reduction of the lattice
connectivity enhances quantum fluctuations in a continuous fashion,
raising the possibility of quantum destruction of magnetic order
{\em before} the percolation threshold is reached. However, recent
studies of homogeneously site- and bond-diluted QHAFs on the square
%v2!
lattice \cite{Katoetal00, Todoetal00A, Sandvik02A, Vajketal02} found
that the magnetic transition takes place exactly at the percolation
threshold, due to the fact that the percolating cluster at threshold
shows LRO \cite{Sandvik02A}. The critical exponents of the
correlation length and of the order parameter, $\nu$ and $\beta$,
are found to take their classical percolation
values\cite{Yasudaetal01A, Sandvik02A}. Therefore, in this case the
magnetic transition is completely dominated by classical
percolation\cite{footnote1}.

Alternative to the above classical percolation picture
is the \emph{quantum percolation} mechanism, recently
demonstrated in a number of model systems
\cite{Yuetal05, Sandvik02B, VajkG02,Sknepneketal04}.
This scenario is based on the fact that spins
involved in locally strongly fluctuating quantum states,
such as dimer singlets and resonating valence bonds (RVBs),
are weakly correlated with the remainder of the system.
In a random network of spins, the local
strongly fluctuating states create
\emph{weak links} with small spin-spin correlations.
If these weak links are part of the backbone
of the percolating cluster, they can prevent the
percolating cluster from developing long-range order.
Therefore, if lattice dilution favors the local
formation of such states, it is possible
to drive the system towards a quantum
disordered state before the percolation threshold
is reached, thus
\emph{decoupling percolation from magnetic ordering}.

In this work we numerically scrutinize this quantum percolation
scenario in the $S=1/2$ QHAF on a square lattice with
\emph{inhomogeneous} bond dilution \cite{footnote2}. The
inhomogeneous character of lattice randomness is the key ingredient
to enhance quantum fluctuations, favoring the local formation of
RVB/dimer-singlet states. Based on large-scale Quantum Monte Carlo
(QMC) simulations, we observe that the classical percolation
scenario is preserved for moderate inhomogeneity, up to a
multicritical point beyond which the magnetic transition deviates
from the percolation threshold. At this point, the quantum
percolation scenario sets in, leading to a novel quantum-disordered
phase. Such a phase has very unconventional features, i.e. a
\emph{finite correlation length} but a \emph{gapless excitation
spectrum}, and a \emph{diverging uniform susceptibility} at zero
temperature with a non-universal exponent. These signatures allow us
to identify this phase with a genuine \emph{two-dimensional quantum
Griffiths phase} \cite{Griffiths69, McCoy69, Hymanetal96}.

This paper is organized as follows.
In Sec. \ref{model} the model is described, and
in Sec. \ref{numerics}, some technical aspects
of the simulations are reviewed.
In Sec. \ref{cluster} we discuss in detail the
fate of antiferromagnetic order on the percolating cluster
upon tuning the inhomogeneity of bond disorder. The complete phase
diagram of the model is discussed in Sec. \ref{phd}.
Sec. \ref{ladder} deals with the emergence of
the quantum-disordered phase in the limit of
randomly coupled ladders, whereas Sec. \ref{GS}
is dedicated to the novel ground state
properties of the quantum-disordered regime.
Conclusions are drawn in Sec. \ref{conclusion}.

\begin{figure}
\begin{center}
\includegraphics[bbllx=0pt,bblly=0pt,bburx=1031pt,bbury=1030pt,%
     width=90mm,angle=0]{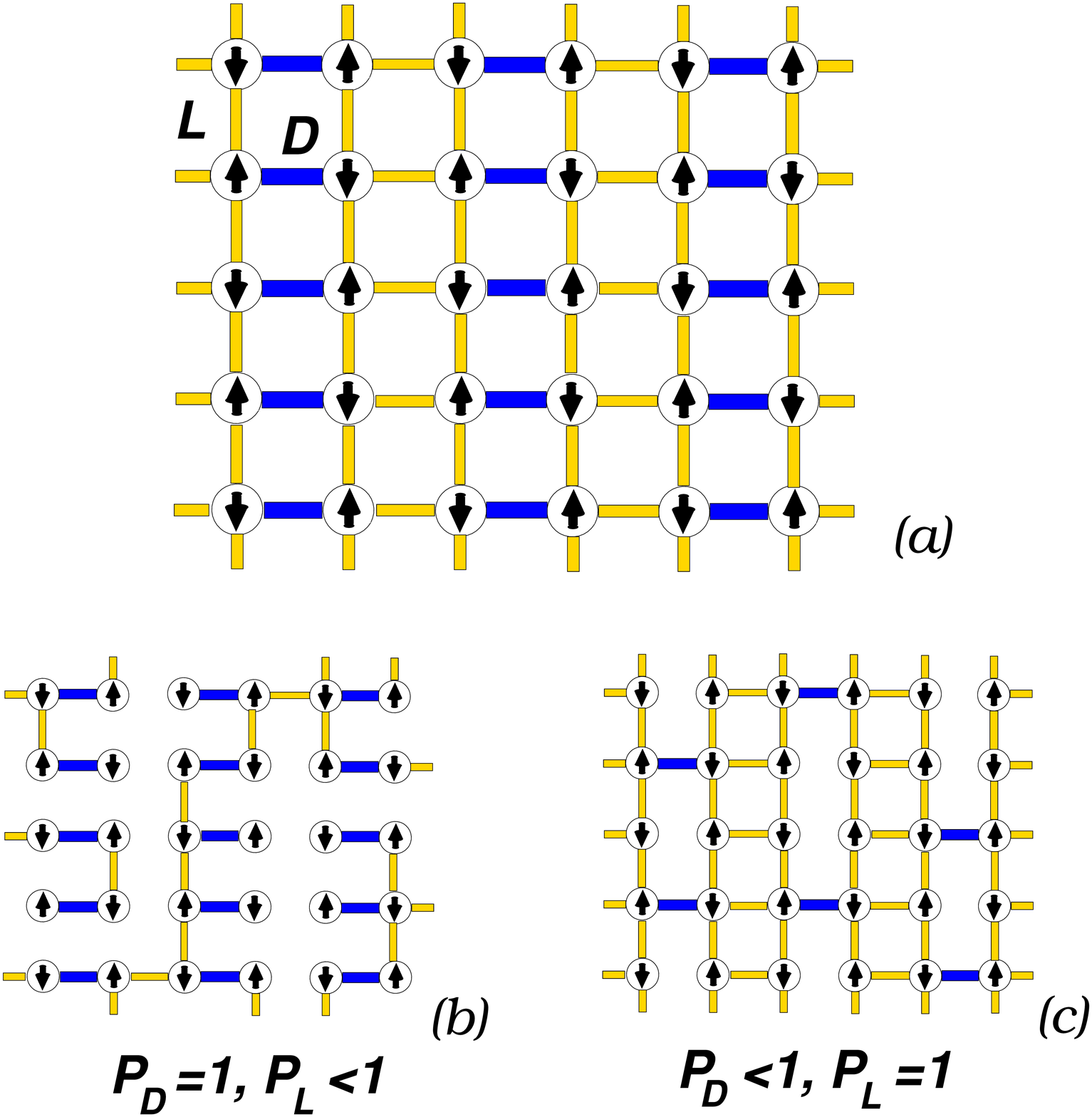}
\caption{\label{f.structure} (color online) (a) Decomposition of the
square lattice into dimer ($D$) and ladder ($L$) bonds; (b) randomly
coupled dimer limit; (c) randomly coupled ladder limit.}
\end{center}
\end{figure}

\section{\label{model}Inhomogeneous bond dilution of the QHAF
on the square lattice}

We investigate the ground state and thermodynamics of the
$S=1/2$ QHAF on a two-dimensional square lattice with
inhomogeneous bond dilution. The Hamiltonian of this system
is given by
\begin{equation}
 {\cal H} = J \sum_{\langle ij \rangle \in D}
 \epsilon^{(ij)}_{_D}~
 {\bm S}_{i}\cdot{\bm S}_{j}+
 J \sum_{\langle lm \rangle \in L}
 \epsilon^{(lm)}_{_L}~
 {\bm S}_{l}\cdot{\bm S}_{m} .
\label{e.hamiltonian}
\end{equation}
The sums run over the
\emph{dimer} ($D$) and \emph{ladder} (inter-dimer)
($L$) bonds, as indicated in Fig. \ref{f.structure}(a).
The exchange couplings are taken to be equal and antiferromagnetic
($J > 0$).  $\epsilon_{\alpha} (\alpha=D,L)$ is a random
variable drawn from a bimodal distribution taking values
$1$ (`on') or $0$ (`off'). The inhomogeneity of this model is
of purely statistical nature, stemming from the different
probabilities of assigning the `on'-state ($\epsilon_i=1$)
to the dimer bonds versus the ladder bonds.
For $D$ bonds this probability is denoted by $p (\epsilon_D=1) =
P_D$, whereas for $L$ bonds it is denoted by $p (\epsilon_L=1) = P_L$.
The special case $P_D=P_L$ represents the previously studied
homogeneous bond-dilution problem \cite{StaufferA94,Sandvik02A}.
In the general inhomogeneous case
$P_D \neq P_L$, two limits are noteworthy (see Fig. \ref{f.structure}(b)-(c)):
the case $P_D = 1$, $P_L \in [0,1]$ corresponds to
a system of randomly coupled dimers, and the opposite case
$P_L = 1$, $P_D \in [0,1]$ corresponds to randomly coupled ladders.
The degree of inhomogeneity can be parametrized by
introducing a variable
\begin{equation}
 \theta = \arctan\left(\frac{1 - P_L}{1 - P_D}\right)
\label{e.theta}
\end{equation}
such that the limiting cases of randomly coupled ladders,
homogeneous bond dilution and randomly coupled dimers
correspond to $\theta=0$, $\pi/4$, and $\pi/2$ respectively.

\begin{figure}
\begin{center}
\includegraphics[
%bbllx=100pt,bblly=20pt,bburx=328pt,bbury=259pt,%
     width=85mm,angle=0]{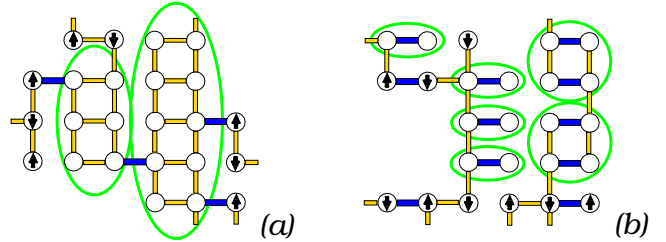}
\caption{\label{f.qlocal} (color online) Inhomogeneous bond dilution
favors the formation of local strongly fluctuating quantum states.
The green ellipses indicate such local dimer-singlet,
plaquette-singlet and ladder-like RVB states: (a) $P_L > P_D$; (b)
$P_D > P_L$. } \vskip -.5cm
\end{center}
\end{figure}

It is evident that the inhomogeneous nature
of bond dilution in the lattice enhances
quantum fluctuations in the magnetic Hamiltonian.
In the strongly inhomogeneous limits $\theta \to 0$, $\pi/2$
the structure of the percolating cluster is geometrically
built from weakly coordinated segments of ladders
($\theta \to 0$),
and weakly coordinated dimers ($\theta \to \pi/2$), \emph{e.g.},
finite necklace-like structures and
finite randomly decorated chains (see Fig. \ref{f.qlocal}).
Both, antiferromagnetic Heisenberg
ladders\cite{DagottoR96} and decorated chains
\cite{Igarashietal95,Moukourietal99,Sierraetal99},
have \emph{quantum-disordered} ground
states with a finite correlation length $\xi_0$.
In antiferromagnetic ladders
the ground state has a RVB nature \cite{Whiteetal94},
and for the necklace it has a dimer-singlet
nature
\cite{Igarashietal95,Moukourietal99}. When segments of such
structures become part of a larger
cluster, as it is the case in the inhomogeneous percolation
model, they locally retain
ground state properties similar to the thermodynamic limit
if their length $l$ is large compared to the correlation
length $\xi_0$. This represents the core mechanism
of non-trivial enhancement of quantum fluctuations
through bond dilution, leading to novel quantum
disordered phases.

\section{\label{numerics} Numerical methods}

The purely geometric problem of inhomogeneous percolation is studied
using a generalized version of a highly efficient classical Monte
Carlo algorithm \cite{NewmanZ02}. In turn, the quantum magnetic
Hamiltonian Eq.~(\ref{e.theta}) is investigated by Stochastic Series
Expansion (SSE) QMC simulations based on the directed-loop algorithm
\cite{SyljuasenS02}. The ground state properties are systematically
probed by efficiently cooling the system down to its physical $T=0$
behavior via a successive doubling of the inverse temperature
$\beta$ (Ref. \onlinecite{Sandvik02A}). This approach is necessary
since in two-dimensional (2D) diluted systems 2D correlations are
often mediated by narrow quasi-one-dimensional links, such that the
temperature scale at which these correlations set in is much lower
than in the clean 2D case.

The scaling analysis of the quantum simulation results
is carried out in two complementary ways \cite{Sandvik02A}.
In the first approach, the lattice size is fixed at
$L \times L$, and the entire system is simulated, including
percolating and non-percolating finite clusters.
In the second approach, a fundamental theoretical tool
is used to ascertain the presence or absence of order in the
system: we focus exclusively on the percolating cluster,
i.e. discarding the finite clusters that cannot carry
long-range order. On a finite $L \times L$ lattice with
periodic boundary conditions the percolating
cluster is identified as the largest cluster, winding
around at least one of the two spatial dimensions.
The advantage of this approach is that a single run of QMC
simulation evaluates the observables on both the full lattice and the
largest (percolating) cluster. At the same time the
largest cluster has
non-negligible size fluctuations (of order $L$),
leading to a significant error in the QMC data.
An alternative approach, eliminating these size fluctuations,
is to grow the percolating cluster freely from an initial seed
up to a fixed size $N_c$ but without any restriction of lattice
boundaries.  Starting from a single occupied bond, one visits
the six neighboring bonds, activating them with probability
$P_D$ ($P_L$) if they are $D$ ($L$) bonds, etc. The procedure
is repeated until no new bonds are generated. The cluster geometry
is accepted only if its final number of sites matches exactly
the required $N_c$. A drawback of this procedure is that
the cluster build-up becomes time consuming for large
$N_c$ and away from the percolation threshold because of
the large rejection rate.

In the full-lattice studies, $L$ values up to 64 were used.
Within the $\beta$-doubling scheme, inverse temperatures as high
as $\beta = 256 L$ have proven to be necessary to observe
the physical $T=0$ behavior. For the studies on
fixed-$N_c$ clusters we considered $N_c$ values up to 2048
and inverse temperatures as high as $\beta = 8N_c$.
For each lattice size and for each point
$\bm{P}$=($P_D$,$P_L$), $10^2$-$10^3$ realizations of the
diluted lattice/percolating cluster
were generated independently to obtain
disorder-averaged observables.

\section{\label{cluster}Magnetism on the percolating cluster}

In this section, QMC results are discussed which address
the evolution of antiferromagnetic order on percolating
clusters upon tuning the nature of bond dilution from
homogeneous ($\theta = \pi/4$) to inhomogeneous
($\theta \to 0$, $\pi/2$).
A fundamental observation \cite{Sandvik02A}
is that the presence or absence of magnetic order in a diluted
system and the nature of the transition from order to disorder
are determined by the magnetic behavior of
the {\emph percolating} cluster. In two dimensions,
only one percolating cluster can exist
in the system. We denote with $\langle M_c \rangle$
the disorder-averaged ($\langle...\rangle$) staggered
magnetization per site
of a percolating cluster, which is estimated by
\begin{equation}
\langle M_c \rangle = \sqrt{\langle m_c^2 \rangle} =
\sqrt{ \left\langle
\frac{3}{N_c^2} {\sum_{ij}}^{(c)} (-1)^{i+j}S_i^{z} S_j^{z}
\right\rangle} .
\end{equation}
Here the summation ${\sum}^{(c)}$ is restricted to
sites contained in the percolating cluster, and
$N_c$ is the total number of these sites.
This is to be contrasted with the overall magnetization
of the diluted lattice with $N=L^2$ sites,
\begin{equation}
\langle M \rangle = \sqrt{\langle m^2 \rangle} =
\sqrt{ \left\langle
\frac{3}{N^2} {\sum_{ij}} (-1)^{i+j}S_i^{z} S_j^{z}
\right\rangle} .
\end{equation}

Given that only the percolating cluster contributes to
long-range order in the thermodynamic limit,
one finds that for $N\to\infty$
\begin{equation}
\langle m^2 \rangle =
\left\langle
\frac{3}{N^2} {\sum_{ij}}^{(c)} (-1)^{i+j}S_i^{z} S_j^{z}
\right\rangle = \left\langle \frac{N_c^2}{N^2}~m_c^2 \right\rangle .
\end{equation}
Moreover, exploiting the self-averaging property
of the size distribution of
the percolating cluster \cite{Sandvik02A},
close to the percolation threshold
one can use
\begin{equation}
\langle M(\bm P) \rangle \approx A |{\bm P} -
{\bm P}^{(cl)}_c|^{\beta}~ M_c(\bm P),
\label{e.magnetization}
\end{equation}
where $\bm P = (P_D,P_L)$ is the control parameter
of the inhomogeneous percolation problem, and ${\bm P}={\bm P}^{(cl)}_c$
defines the classical critical line of percolation thresholds in the
$(P_D,P_L)$ space. In Eq.~(\ref{e.magnetization})
we have used the critical behavior of the so-called
network strength $\langle N_c/N \rangle \approx
A |{\bm P} - {\bm P}^{(cl)}_c|^{\beta}$,
where $A$ is a constant amplitude, and $\beta=5/36$ is
the critical exponent of classical percolation
\cite{StaufferA94}.

From Eq.~(\ref{e.magnetization}) it is evident
that, if $M_c(\bm P^{(cl)}_c) \neq 0$, the dilution-driven
transition from magnetic order to disorder coincides
with the classical percolation transition, both
in terms of its location and of its critical exponents.
Therefore only the absence of antiferromagnetic
order on the percolating cluster can lead to a
decoupling between the magnetic transition and
the geometric percolation threshold.

\begin{figure}
\begin{center}
\includegraphics[
%bbllx=100pt,bblly=20pt,bburx=328pt,bbury=259pt,%
     width=75mm,angle=0]{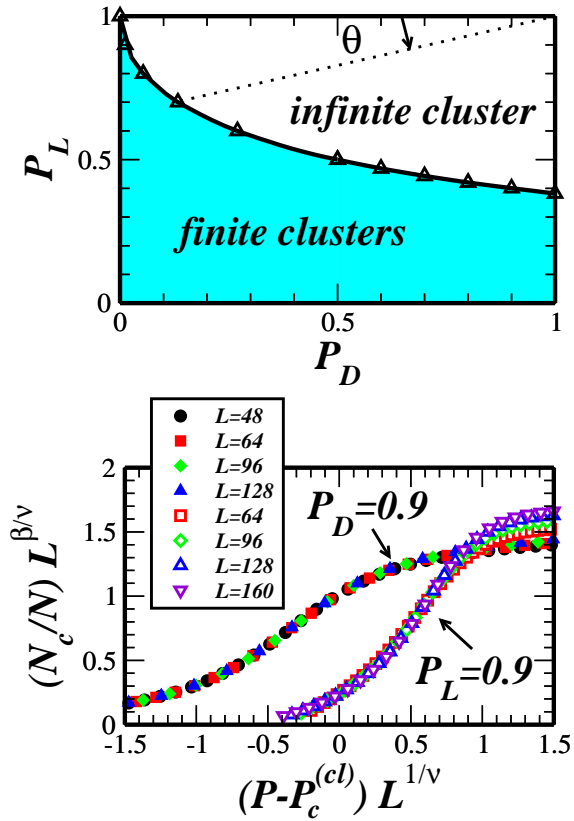}
\caption{\label{f.percolation} (color online) \emph{Upper panel}:
classical phase diagram for the inhomogeneous bond-percolation
problem on the square lattice. The angle $\theta$ parametrizing the
critical curve is indicated. \emph{Lower panel}: scaling plots for
the network strength $N_c/N$ in the two strongly inhomogeneous cases
$P_D=0.9$ and $P_L=0.9$. An excellent data collapse is realized,
using the 2D percolation exponents $\beta=5/36$ and $\nu=4/3$, and
with $P_{L,c}^{(cl)}=0.4007$ ($P_D=0.9$) and $P_{D,c}^{(cl)}=0.0107$
($P_L=0.9$).} \vskip -.5cm
\end{center}
\end{figure}

The critical curve ${\bm P}={\bm P}^{(cl)}_c$, determined from
classical Monte Carlo simulations, is shown in Fig.
\ref{f.percolation}. Despite the inhomogeneous nature of percolation
for $P_L \neq P_D$, excellent scaling of the classical MC results is
observed using the 2D percolation exponents \cite{StaufferA94}. This
demonstrates that the universality class remains unchanged along the
critical curve.

\begin{figure}
\begin{center}
\includegraphics[
%bbllx=100pt,bblly=20pt,bburx=328pt,bbury=259pt,%
     width=85mm,angle=0]{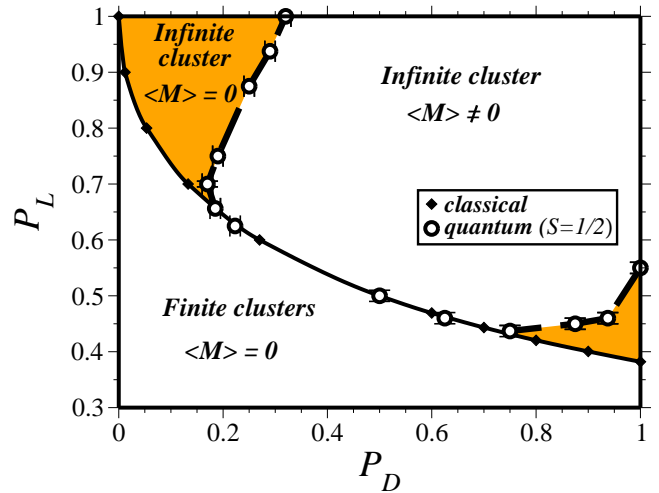}
\caption{\label{f.phasediagr} (color online) Phase diagram of the
spin-$1/2$ QHAF on the inhomogeneously bond-diluted square lattice.}
\vskip -.5cm
\end{center}
\end{figure}

Starting from the classical Monte Carlo result,
the evolution of the order parameter can be tracked on the
percolating cluster $\langle M_c({\bm P}^{(cl)}_c) \rangle$ along the critical
line ${\bm P} = {\bm P}^{(cl)}_c$ to monitor the effect of
quantum fluctuations enhanced by inhomogeneity.
To this end, we perform a scaling study of the staggered
magnetization on percolating clusters for fixed size
$N_c$, with $N_c$ up to 2048 and $\beta = 8 N_c$.
The obtained data are extrapolated to
the thermodynamic limit by a three-parameter polynomial
fit $\langle m_c^2 (N_c) \rangle =
\langle m_c^2 (\infty)\rangle + a~ N_c^{-1/2} + b~ N_c$.

At the homogeneous bond percolation threshold
$P_D=P_L=0.5$ the percolating cluster is found to
have antiferromagnetic
long-range order, in agreement with
Ref. \onlinecite{Sandvik02A}. However, the finite value of
the order parameter is strongly reduced as the
inhomogeneity is turned on, both in the "dimer"
($\theta\to \pi/2$) and
the "ladder" ($\theta\to 0$) directions, until
it  \emph{vanishes} at non-trivial
$\theta$ values, $\theta_L \approx 0.34$
in the ladder limit and $\theta_D \approx 1.23$
in the dimer limit. This means that
inhomogeneous percolation exhibits
\emph{non-linear} quantum fluctuations that
are able to destroy the
antiferromagnetic long-range order, in contrast to
the homogeneous percolation case\cite{Sandvik02A}
with a renormalized classical ground state.

According to
Eq. (\ref{e.magnetization}) this must be reflected in
a profound change of the critical properties,
since the vanishing of the
staggered magnetization
on the percolating cluster decouples the magnetic
transition from the percolation threshold.
Beyond the two critical values $\theta_L$ and
$\theta_D$, a higher concentration of
bonds will be required for the system to
magnetically order than for the lattice to
percolate, thus opening up an
intermediate phase with a novel quantum-disordered
ground state.

\begin{figure}
\begin{center}
\includegraphics[
%bbllx=100pt,bblly=20pt,bburx=328pt,bbury=259pt,%
     width=85mm,angle=0]{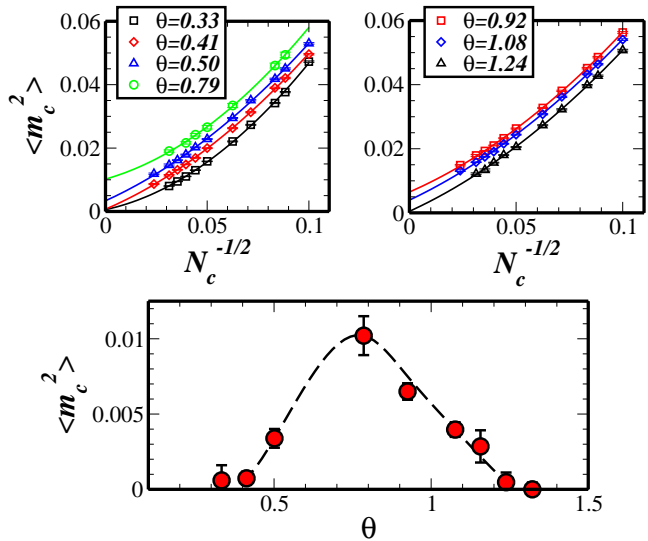}
\caption{\label{f.mag.Nc} (color online) \emph{Upper panel}: scaling
of the disorder-averaged squared staggered magnetization on the
percolating cluster with fixed size $N_c$. The different curves
correspond to various points along the classical percolation
transition. The continuous lines represent quadratic fits.
\emph{Lower panel}:  extrapolated thermodynamic values of the
staggered magnetization from the upper panel as a function of the
inhomogeneity parameter $\theta$. The dashed line is a guide to the
eye.} \vskip -.5cm
\end{center}
\end{figure}

\section{\label{phd}Phase diagram}

In this section, QMC results are discussed for the magnetic phase
diagram in the ($P_D$,$P_L$) plane. The magnetic transitions are
located via scaling of the spin-spin correlation length $\xi$,
extracted from the structure factor through the second moment
estimator \cite{Cooperetal82}. The inhomogeneity of bond dilution
breaks the discrete rotation symmetry of the lattice, such that two
distinct correlation lengths $\xi_x$ and $\xi_y$ need to be
considered along the $x$ and $y$ lattice directions. At critical
points separating 2D long-range order from disordered phases,
\emph{both} correlation lengths must scale linearly with the lattice
size, $\xi_x , \xi_y \sim L$, regardless of the universality class
of the transition. In locating the magnetic transitions we have
verified that this condition is satisfied. This leads to the
conclusion that, despite its inhomogeneity, the magnetic system has
unique well-defined phase transitions.

The resulting phase diagram is shown in Fig. \ref{f.phasediagr}. The
magnetic transition line coincides with the percolation line for
moderate inhomogeneity. As discussed in the previous section, in
this region of the phase diagram the percolating cluster at the
percolation threshold is antiferromagnetically ordered. Therefore
not only the magnetic transition and the geometric transition
coincide, but also the critical exponents of the magnetic transition
are those of 2D percolation \cite{Yuetal05}, according to
Eq.~(\ref{e.magnetization}).

For strong inhomogeneity. the scenario
of a classical percolation transition in the
magnetic Hamiltonian is precluded by
non-linear quantum fluctuations. In fact,
two multicritical
points occur, in the ladder and in the dimer
direction, beyond which the magnetic transition
decouples from the percolation threshold. The location
of these multicritical points is
quantitatively consistent with the vanishing
of antiferromagnetic order on the percolating
cluster at two critical values of
the inhomogeneity parameter $\theta$ (Fig. \ref{f.mag.Nc}).

\begin{figure}[h]
\begin{center}
\includegraphics[width=80mm]{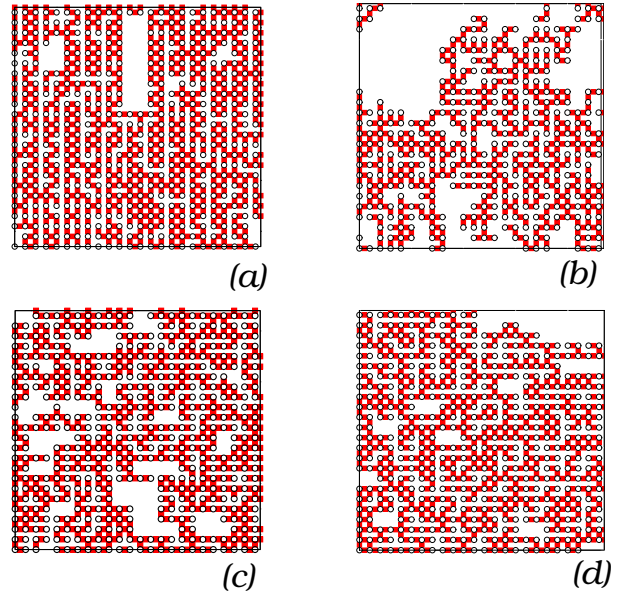}
\caption{\label{f.structures} (color online) Percolating
 clusters for the bond-diluted square lattice with size
 $24\times24$ and periodic boundary conditions:
 (a) $P_D=0.15$, $P_L=0.8$;
 (b) $P_D=0.5$, $P_L=0.5$ (homogeneous case);
 (c) $P_D=0.9$, $P_L=0.45$;
 (d) $P_D=1.0$, $P_L=0.44$ (randomly coupled dimers).}
\end{center}
\end{figure}

 Beyond the two multicritical points, intermediate
quantum phases appear, in which antiferromagnetic order is absent on
the percolating cluster both at and away from the percolation
threshold. These regions represent novel quantum-disordered phases
on an infinite family of 2D percolated random lattices.
Representative structures of this family are shown in Fig.
\ref{f.structures}, together with the homogeneous percolating
cluster. It is remarkable that the homogeneous percolating cluster
(Fig. \ref{f.structures}(b)) has long-range antiferromagnetic order
in the thermodynamic limit, whereas the other percolating clusters
in Fig. \ref{f.structures}(a),(c),(d) do not. In fact, the latter
structures have a significantly higher average number of bonds per
site than the homogeneous percolating cluster, which has a fractal
dimension\cite{StaufferA94} lower than 2 ($D_f = 91/48$). A detailed
discussion of the nature of the quantum-disordered phase will be
given in Sec. \ref{GS}.

 The evolution of the critical exponents
upon tuning inhomogeneity has been investigated
in Ref. \onlinecite{Yuetal05}. Surprisingly,
the two multicritical points do not mark an evident
discontinuity in the universality class of
the model, and the classical percolation exponents
appear to persist also for intermediate inhomogeneity
beyond the multicritical points, changing
to radically different values only in the extreme
inhomogeneous limits. Therefore the magnetic \emph{quantum}
phase transitions close to the multicritical points
appear to retain a percolative character.
This result can be understood within the following
argument.
To destroy the long-range nature of a
random network it is sufficient to cut a few
links on its backbone. This implies that
the non-linear quantum fluctuations
leading to quantum-disorder on the percolating
cluster need only have \emph{short-wavelength}
components, in contrast to conventional quantum phase
transitions in translationally invariant lattices.
Indeed, strongly fluctuating states
appearing on links made of segmented ladders or
decorated chains (see Sec. \ref{model})
have a markedly local effect on magnetic
correlations, i.e. they
magnetically decouple the portions of the
percolating cluster connected through that link.
This effect is equivalent
to geometrically removing that link, thus
leaving two portions of the cluster
disconnected \cite{Yuetal05}.
If weak links occur on the backbone
of the percolating cluster, this mechanism is
sufficient to lead to quantum disorder. Only
quantum fluctuations with wavelengths comparable to
the length of the weak link are required.
To reestablish magnetic correlations between the two
portions of the percolating cluster disconnected
by quantum fluctuations, it is then necessary to
add more bonds to the cluster in order to
find an alternative path for magnetic correlations
to spread over the cluster. Since this process is
of geometrical character it
endows the quantum phase transition with
a percolative nature.

 The above picture of a percolative quantum phase
transition breaks down for strong inhomogeneity.
In this limit, the building blocks of the percolating
cluster are the strongly quantum fluctuating substructures
depicted in Fig.~\ref{f.qlocal}, and hence the quantum
phase transition is driven by the competition between
the energy scale of the gap above the ground state
of such substructures and the energy scale of the
spatially random couplings between the substructures.
In a real-space picture, the competition occurs between
the finite correlation length of the substructures
and the characteristic length set by the spacing
between the spatially random couplings. This competition
will be discussed in detail in the next section.
The quantum phase transition is completely disconnected from
classical percolation of the lattice, and it is
reasonable to expect for it to be in a different
universality class, as concluded in
Ref.~\onlinecite{Yuetal05}.

\section{\label{ladder}Criticality near the ladder limit}

So far, the onset of the quantum disordered phases was discussed,
approaching them from the homogeneous limit of
Eq.~(\ref{e.hamiltonian}). A complementary understanding of the
novel quantum phase and the associated quantum phase transitions can
be gained by starting from the extreme inhomogeneous limit of
randomly coupled ladders ($\theta=0$). This limit lends itself to a
very simple analysis in terms of the properties of the single
spin-$1/2$ antiferromagnetic Heisenberg ladder.

In the limit of $P_L = 1$, the system is composed of an array of
two-leg ladders randomly coupled by $D$ bonds. The probability
of activation of a dimer bond is reflected
in the average distance $\langle r \rangle$ of two
$D$ bonds along a column separating two neighboring ladders, namely
$P_D = 1/\langle r \rangle$.
A single $D$ bond is sufficient to connect two ladders
geometrically. In other words,  $
\langle r \rangle = L$ is sufficient to connect all ladder subsystems
on a $L \times L$ lattice. Taking the limit
$L \rightarrow \infty$, one obtains immediately $P_D \rightarrow 0$.
Thus geometrically the system percolates at any
infinitesimal concentration $P_D > 0$ and the percolating
cluster is the entire lattice itself.

\begin{figure}[h]
\begin{center}
\includegraphics[
%bbllx=-20pt,bblly=-20pt,bburx=600pt,bbury=466pt,%
     width=85mm,angle=0]{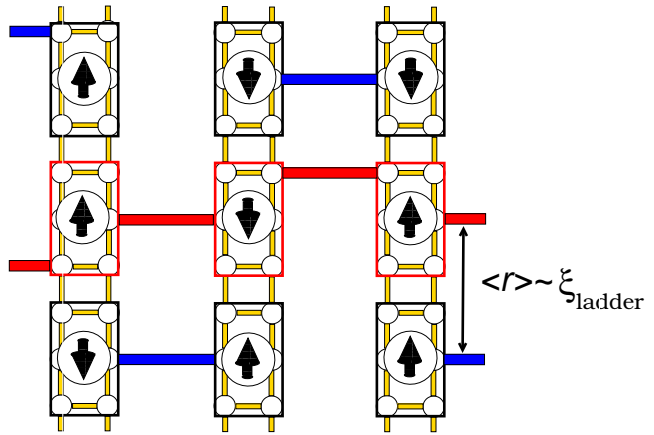}
\caption{\label{f.RGladder} (color online) Cartoon of the ordering
mechanism in randomly coupled ladders. When $P_D = 1/\langle
r\rangle \approx 1/\xi_{\text{ladder}}$ (see text) percolating
strings of block spins (rectangles) appear in the system. The
divergence of the correlation length along such strings drives the
onset of long-range correlations between the strings and ultimately
of 2D long-range order.} \vskip -.5cm
\end{center}
\end{figure}

However the system does not develop antiferromagnetic order until a
finite value $P_D \approx 0.32$ is reached. This is not surprising
since each two-leg ladder has a quantum-disordered ground state with
a finite excitation gap. This gap has to be overcome before the
N\'eel ordered state can become the ground state. A more
quantitative argument can be formulated based on the fact that each
isolated ladder has a finite correlation length \cite{Whiteetal94}
$\xi_{\text{ladder}} \approx 3.19$. One can imagine a
renormalization group transformation that creates effective block
spins from all the spins within a correlation volume in such a way
that all block spins are perfectly uncorrelated with each other (see
Fig. \ref{f.RGladder}). When turning on the $D$ bonds, it becomes
evident that, to establish long-range correlations in the system, it
is necessary to have a string of block spins percolating from one
side to the other of the system. Given that all block spins are
independent within the same ladder, this is only possible if, on
average, each block spin is connected by a $D$ bond to both its
right and left neighboring block spin. This requires an average
spacing of the $D$ bonds of $\langle r \rangle \approx
\xi_{\text{ladder}}$ on each column, which in turn leads to the
following estimate of the critical concentration of dimer bonds
\begin{equation}
P_{D,c} = \frac{1}{\xi_{\text{ladder}}}. \label{e.ladder}
\end{equation}
Despite the simplicity of the argument, this
estimate is surprisingly good:
$P_{D,c} \approx 0.313$ to be compared with the
QMC result $P_{D,c} = 0.32(1)$.

\begin{figure}[h]
\begin{center}
\includegraphics[
%bbllx=-30pt,bblly=-30pt,bburx=750pt,bbury=700pt,%
     width=70mm,angle=0]{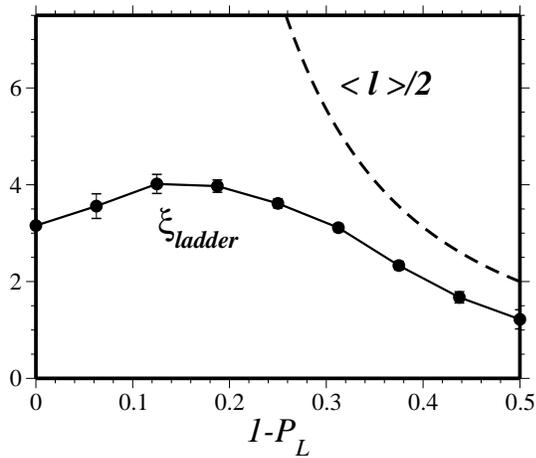}
\caption{\label{f.xiladder} Correlation length of the
bond-diluted $128\times 2$ ladder compared with the average ladder
length $\langle l \rangle$. The data for $\xi_{\rm ladder}$ are
taken at $\beta = 2048$; comparison with data at $\beta=4096$ (not
reported)
shows no significant deviation.} %\vskip -.5cm
\end{center}
\end{figure}

 Interestingly, this argument can be extended to the
case of randomly coupled \emph{weakly diluted} ladders,
namely for $P_L \lesssim 1$.
The correlation length of the ladder remains the
dominant length scale of the problem, as long as
it is much smaller than
the new length scale introduced by the dilution
\begin{equation}
\langle l\rangle = \frac{1}{(1-P_L)^2},
\end{equation}
corresponding to
the average length of the ladder segments after dilution.

\begin{figure}[h]
\begin{center}
\null\hspace{0.7cm}\includegraphics[
%bbllx=30pt,bblly=20pt,bburx=195pt,bbury=250pt,%
     width=70mm,angle=0]{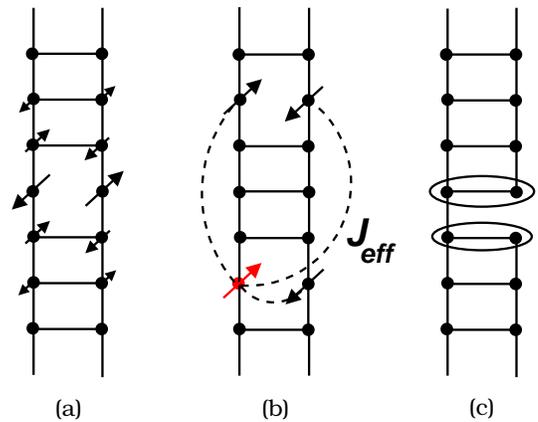}
\caption{\label{f.diluted-ladder} (color online) Dominant effects of
bond dilution on a two-leg ladder: (a) local antiferromagnetic
modulation due to a missing rung bond; (b) effective couplings
between spins missing their rung bond; (c) enhancement of the
singlet component (indicated by an ellipse) of the two rung dimers
adjacent to a missing leg bond.}
\end{center}
\end{figure}

To test to which extent the above argument is
applicable
we performed SSE-QMC simulations on a \emph{single}
bond-diluted two-leg ladder to determine
the evolution of the correlation length upon
dilution. The results are shown in Fig. \ref{f.xiladder}.
For low doping concentrations, we observe that
the average correlation length of the ladder
increases moderately. For higher concentrations
of missing bonds, the average ladder length $\langle l \rangle$
becomes comparable to the correlation length,
which then
crosses over to a decreasing behavior simply
reflecting that of $\langle l \rangle$.

We now focus on the increasing tendency
for small dilution. This behavior reflects
various competing mechanisms. The dominant effects having the
highest probabilities (${\cal O}(P_L)$ and ${\cal O}(P_L^2)$)
are: \\
(a) The removal of a rung bond
results locally in two uncoupled
$S=1/2$ free moments which introduce
a local staggered modulation of the
adjacent spins \cite{Fukuyamaetal96,Mikeskaetal97} within
a correlation volume  $\sim \xi_{\text{ladder}}$
(Fig. \ref{f.diluted-ladder}(a)). This
leads to a local increase in the antiferromagnetic
correlations along the legs; \\
(b) If two rung bonds are removed sufficiently close
to each other, the induced
local antiferromagnetic modulations
of the ladder ground state
``lock in phase", leading
to an effective coupling between the
free moments, exponentially decaying with
the distance in the limit of low doping
\cite{SigristF96,Mikeskaetal97},
\begin{equation}
J_{\rm eff} \sim (-1)^{i-j} \exp{[-|i-j|/\xi_{\text{ladder}}]}.
\end{equation}
These long-range interactions also lead
to an increase of the correlation length in the
system through an
\emph{order-by-disorder} mechanism
(Fig. \ref{f.diluted-ladder}(b)); \\
(c) The removal of one leg bond,
on the contrary, leads to a local enhancement
of the rung-singlet component of the state
of the two adjacent rungs
(Fig. \ref{f.diluted-ladder}(c)).
This mechanism weakens the correlations
along the two legs.\\

Effect (c) has twice the
probability compared to the effects
(a) and (b) because there are twice as many leg bonds as
rung bonds.
The non-trivial competition between correlation enhancement
(a)-(b) and correlation suppression
(c) has the combined effect that in
\emph{bond-diluted} ladders, the correlation length
always remains \emph{finite} down to zero temperature.
This is to be contrasted
with the case of \emph{site-diluted} ladders, in which the
correlation length diverges algebraically with decreasing
temperature, $\xi \sim T^{-0.4}$, as reported
in Ref. \onlinecite{GrevenB98}.
(Obviously, even in the site-diluted case
the average correlation length is eventually
upper-bounded by the average length of the ladder
segments). It is interesting to note that for the
case of site dilution, correlation-suppressing
effects are \emph{not} present beside the simple
fragmentation of the ladder.

\begin{figure}[h]
\begin{center}
\includegraphics[
%bbllx=30pt,bblly=20pt,bburx=195pt,bbury=250pt,%
     width=80mm,angle=0]{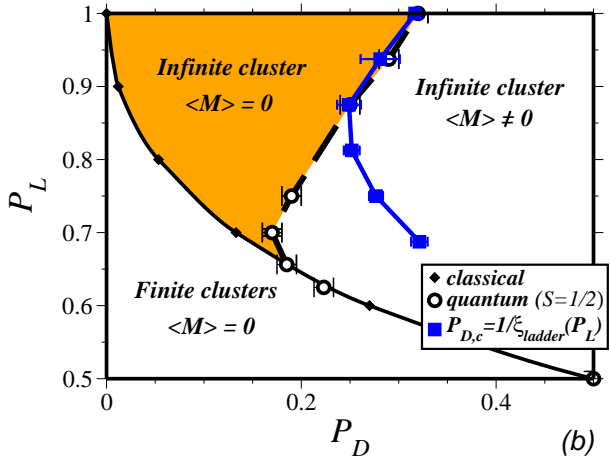}
\caption{\label{f.PhDladder} (color online) Phase diagram of the
QHAF on the inhomogeneously bond-diluted square lattice (see also
Fig.~ \ref{f.phasediagr}) close to the ladder limit and compared
with the single-ladder prediction $P_{D,c} =
1/\xi_{\text{ladder}}(P_L)$ (see text). }
\end{center}
\end{figure}

 The absence of diverging correlations in bond-diluted
ladders is crucial for
the occurrence of an extended quantum-disordered
phase. The finite
correlation length of a single ladder
for $P_L \lesssim 1$ allows us to repeat the
RG argument sketched in Fig. \ref{f.RGladder}
for the case of bond-diluted ladders.
One needs a \emph{finite} and \emph{non-classical}
concentration
of dimer bonds $P_{D,c}$ to drive the system
into a N\'eel ordered phase, directly related
to the \emph{quantum} correlation length of the
system $\xi_{\text{ladder}}$, as long as it is significantly
lower than the \emph{classical} length $\langle l \rangle$:
\begin{equation}
P_{D,c}(P_L) \approx \frac{1}{\xi_{\text{ladder}}(P_L)}~.
\label{f.Pcladder}
\end{equation}
The above estimate is a theoretical
prediction for the quantum-critical curve
${\bm P}_c = (P_{D,c},P_{L,c})$, which can
be directly compared with the QMC results,
shown in Fig. \ref{f.PhDladder}.  The range of
quantitative validity of Eq. \ref{f.Pcladder} is
surprisingly large, extending down to $P_L \approx 0.875$,
which is interestingly also the value where
the quantum correlation length $\xi_{\text{ladder}}(P_L)$
starts to cross over towards the decreasing
classical behavior (Fig. \ref{f.xiladder}).
At this point the ground state of the
ladder segments is strongly altered by the
presence of the dimer bonds, and hence the
argument leading to Eq.~(\ref{f.Pcladder}) breaks
down.

\section{\label{GS} The quantum-disordered phase:
correlations and Griffiths-McCoy singularities}

 In this section, we discuss the
nature of the quantum-disordered phase.
Ground-state properties
and thermodynamic observables are considered
which reveal the nature of
the low-lying excitation spectrum.
A fundamental aspect of this phase, as it is
typical for disordered systems, are
large variations in the local properties of
the system, reflecting the local geometric
structure defined by disorder. In the previous
sections, we have discussed how the presence
of segments of ladders or decorated chains
leads to the local formation of RVB/dimer-singlet
states. At the same time, bond dilution
of ladders leads to the appearence of $S=1/2$
local degrees of freedom, effectively interacting
across the strongly quantum fluctuating regions.
The coexistence of such different subsets of
spins within the same system raises the question
of what the global nature of the ground state
really is.

\subsection{Short-range correlations}

\begin{figure}[h]
\begin{center}
\includegraphics[
%bbllx=30pt,bblly=20pt,bburx=195pt,bbury=250pt,%
     width=80mm,angle=0]{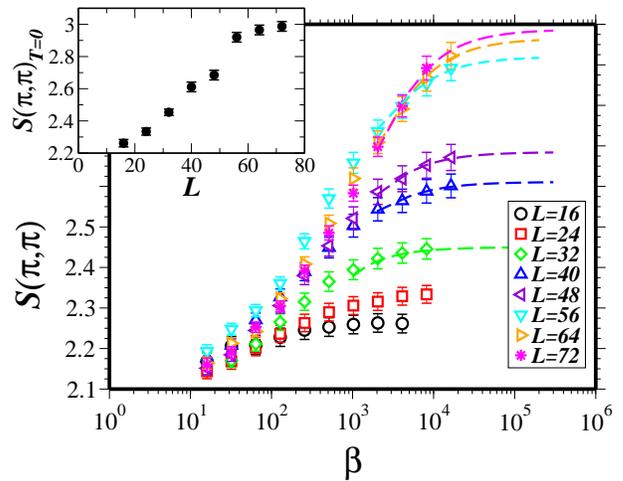}
\caption{\label{f.Spp} (color online) Temperature and finite-size
scaling of the static structure factor $S(\pi,\pi)$ for a
representative point in the quantum-disordered regime ($P_D=0.15$,
$P_L=0.85$). Dashed lines are a quartic polynomial fit in
$T=\beta^{-1}$ [see Eq.~(\ref{e.fit})] to extrapolate to the
$\beta\to\infty$ limit (when required). Inset: $\beta\to\infty$
extrapolated values of the static structure factor plotted
\emph{vs.} system size, clearly showing saturation in the
thermodynamic limit.}
\end{center}
\end{figure}

\begin{figure}[h]
\begin{center}
\includegraphics[
%bbllx=30pt,bblly=20pt,bburx=195pt,bbury=250pt,%
     width=90mm,angle=0]{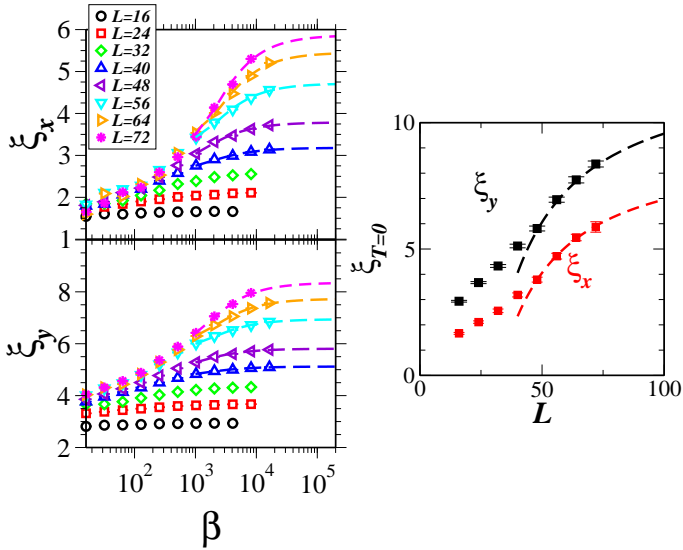}
\caption{\label{f.xi} (color online) \emph{Left panel}: temperature
and finite-size scaling of the correlation length along the two
lattice directions ($\xi_x$ and $\xi_y$) for a representative point
in the quantum-disordered regime ($P_D=0.15$, $P_L=0.85$). Dashed
lines are a quartic polynomial fit in $T=\beta^{-1}$ to extrapolate
to $\beta\to\infty$ (when required). \emph{Right panel}:
$\beta\to\infty$ extrapolated values of the correlation length(s)
\emph{vs.} system size; the dashed lines are polynomial fits up to
second order in $L^{-1}$ on the points for $L\geq 48$ to compensate
for the finite-size effects.}
\end{center}
\end{figure}

\begin{figure}[h]
\begin{center}
\includegraphics[
%bbllx=30pt,bblly=20pt,bburx=195pt,bbury=250pt,%
     width=70mm,angle=0]{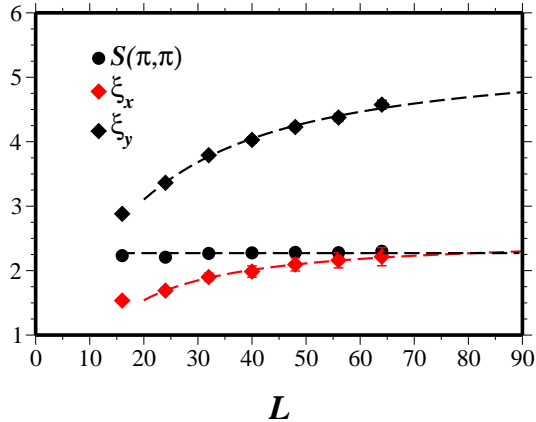}
\caption{\label{f.xi2} (color online) Finite-size scaling of the
$T=0$ static structure factor and correlation length(s) for a point
in the quantum-disordered regime ($P_D=0.15$, $P_L=0.9$) closer to
the ladder limit. The dashed lines associated with the correlation
length data are quadratic fits $\xi(L) = \xi(\infty) + a/L +
b/L^2$.}
\end{center}
\end{figure}

 We first characterize
the ground state in terms of global correlation
properties. The absence of magnetic order in this
phase has already been discussed in Sec. \ref{cluster}.
Fig. \ref{f.Spp} shows the staggered structure factor
$S(\pi,\pi)$ for a representative point
($P_D=0.15$, $P_D=0.85$) in the quantum
disordered phase on the "ladder" side of the
phase diagram in Fig. \ref{f.phasediagr}. The finite-temperature
data for $S(\pi,\pi)$, obtained using $\beta$-doubling,
are extrapolated to the limit $T\to 0$
by accounting for power-law temperature corrections up
to fourth order:
\begin{equation}
S(\pi,\pi;T) = S(\pi,\pi)_{_{T=0}} + \sum_{i=1}^{4} \frac{a_i}{\beta^i}
\label{e.fit}
\end{equation}
where the $a_i$'s are fitting coefficients.
The extrapolated $T=0$ values of $S(\pi,\pi)$
are shown in the inset of Fig. \ref{f.Spp}
as a function of system size.
They clearly display saturation for $L\to\infty$,
proving not only that the system is disordered,
but also that it has \emph{finite-range correlations}.

This fact is further reflected in the correlation length
$\xi_{x(y)}$ along the two lattice dimensions, shown in Fig.
\ref{f.xi}. A fitting procedure similar to Eq.~(\ref{e.fit}) is used
to eliminate polynomial finite-temperature correction. The
$\xi_{T=0}$ values so obtained (right panel of Fig. \ref{f.xi})
suggest convergence towards a finite value for $L \to \infty$,
although the correlation length is large for the particular point
($P_D=0.15$, $P_D=0.85$), and, even for the largest considered size
($L=72$), $\xi$ does not display full saturation. A polynomial fit
$\xi(L) = \xi_{\infty} + a/L + b/L^2 $ for $L\geq 48$ yields
saturation values $\xi_{x,\infty} = 9(1)$ and $\xi_{y,\infty} =
12.5(5)$, resulting in a considerable correlation volume of
$\sim 100$ sites. It is important to keep in mind that the finite
correlation length in the quantum disordered system varies strongly
with the degree of inhomogeneity, and it can become as small as $\xi\sim
3$ in the limit of randomly coupled ladders. For instance,
Fig.~\ref{f.xi2} shows the finite-size scaling of the static
structure factor and correlation lengths for a point closer to the
ladder limit, $P_D = 0.15$ and $P_L=0.9$. For this point one
observes excellent temperature saturation of the data up to $L=64$
for $\beta \leq 16384$ without the need of any fitting procedure.
Moreover, the zero-temperature data reported in Fig.~\ref{f.xi2}
show a much better saturating behavior within the system sizes
considered, leading to the asymptotic values $\xi_{x,\infty} =
2.5(1)$ and $\xi_{y,\infty} = 5.4(2)$.

\begin{figure}[h!]
\begin{center}
\includegraphics[
 %bbllx=50pt,bblly=-20pt,bburx=400pt,bbury=400pt,%
  width=68mm,angle=0]{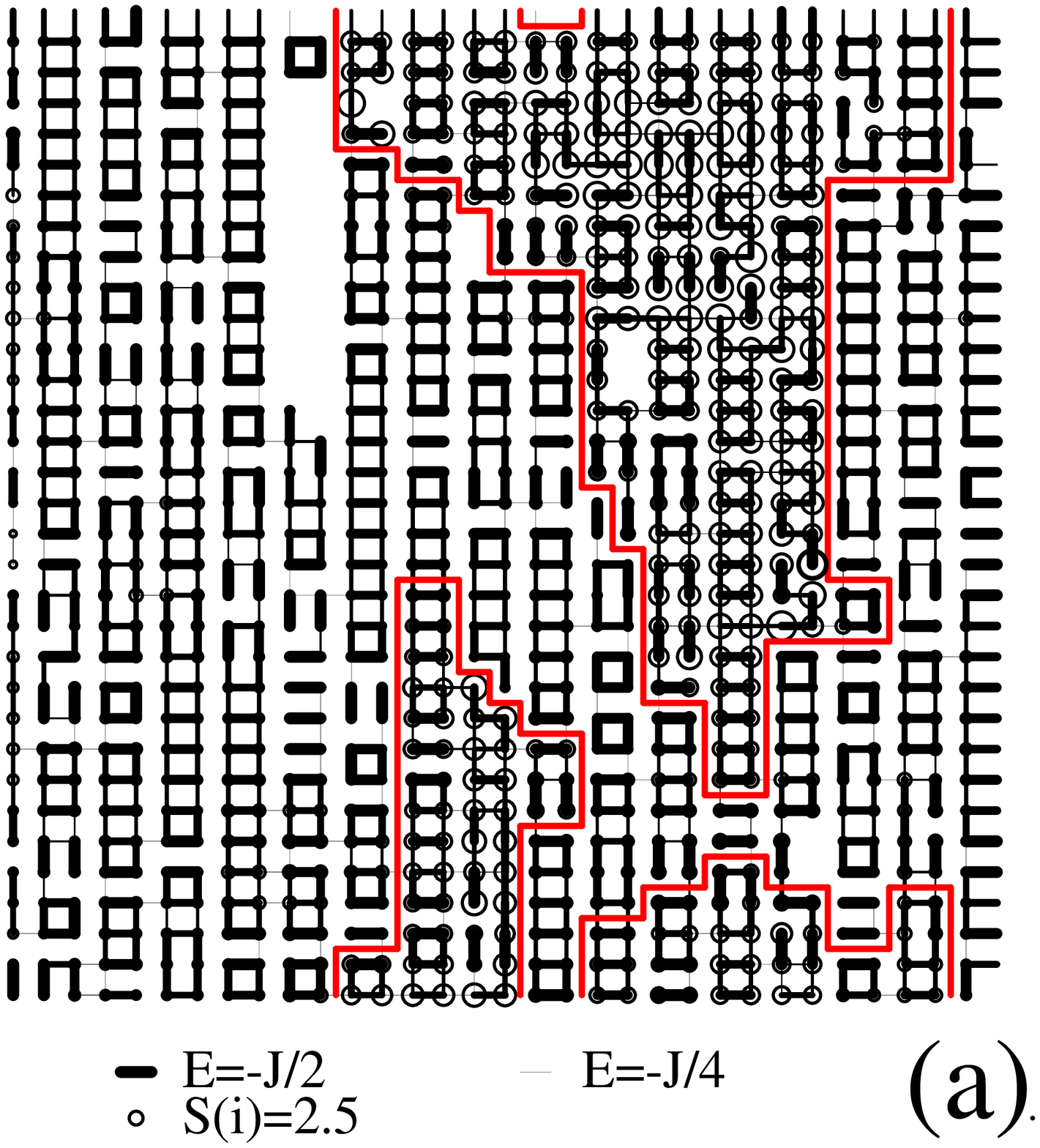}
  \vskip .5cm
\includegraphics[
%bbllx=50pt,bblly=-20pt,bburx=400pt,bbury=400pt,%
     width=68mm,angle=0]{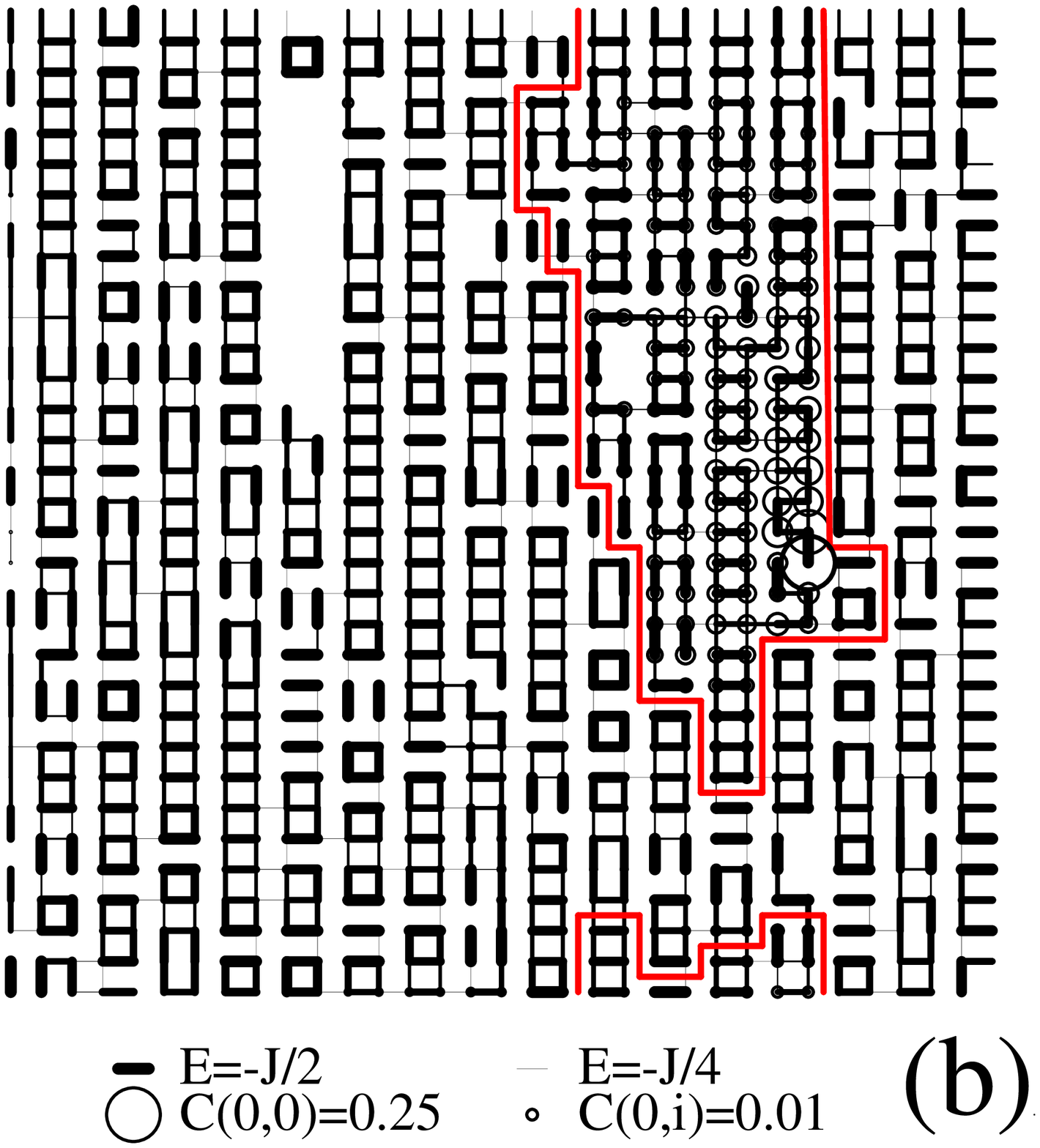}
 \caption{\label{f.realspace} (color online) Real-space images of the
 bond energy (both panels), local static
 staggered structure factor $S(i)$ (see text) (a) and of the static staggered
correlations $C(0,i)$ between a reference point on a dangling bond
and the remainder of the system (b), for the percolating
cluster on a $32\times32$ sample with $P_D=0.15$ and $P_L=0.85$
at $\beta = 8192$.
The thickness of each bond is proportional to its energy,
whereas the radius of the dots in (a) and (b) is proportional
to the local value of $S(i)$ and $C(0,i)$.
In (a) red boundaries highlight the regions
where local correlations exceed the average value,
namely where $S(i) \gtrsim S(\pi,\pi)$. In (b)
the red boundaries mark the correlation volume beyond which
$C(0,i) \leq 8 \times 10^{-3}$, corresponding to a distance
(in units of the correlation length)
$|r_i|/\xi \approx \ln[C(0,0)/C(0,i)]\approx 3.4$~.}
\end{center}
\end{figure}

 A clear picture of the short-range nature of
correlations also emerges from real-space
images of
\begin{itemize}
\item the local staggered structure factor,
\begin{equation}
S(\pi,\pi;i) = \sum_{j\neq i} (-1)^{i+j} \langle S^{z}_i S^{z}_j \rangle_{_0} ;
\label{e.Sloc}
\end{equation}
\item the correlation function
$C(0,i) = \langle S^{z}_0 S^{z}_i \rangle_{_0}$
between a reference site and all the other spins;
\item the energy of each bond
$E_{\langle ij \rangle} = J\langle {\bm S}_i\cdot {\bm S}_j \rangle_{_0}$,
\end{itemize}
all shown in Fig. \ref{f.realspace} for a given sample
corresponding to a representative point of the quantum-disordered
phase on the ``ladder side''. Here the symbol
$\langle...\rangle_{_0}$ indicates the expectation value
associated with the ground state of the specific
sample considered, not to be confused with the disorder
average $\langle...\rangle$.

Starting with the bond energy, we observe that
quantum fluctuations introduce a wide range
of variability in its values, energetically promoting
weakly correlated dimers or
quadrumers and rung bonds on ladder segments,
at the expenses of the energy of the adjacent bonds
(due to the fact that a $S=1/2$ particle can form a
singlet with only one other $S=1/2$ particle).
In particular the rare $D$ bonds have
in general the weakest energies, due to the
strong tendency of the energy to be minimized on the
ladder segments. This leads to
a clear quantum suppression of correlations in the
direction transverse to the ladders.

  The local staggered structure factor of Eq.~(\ref{e.Sloc})
 measures the ``effective correlation volume"
 (integral of the antiferromagnetic correlation function)
 around each spin, thus identifying the spins
 that are most strongly correlated with the remainder of the system.
 In Fig. \ref{f.realspace}(a) we observe that only a portion
 of the spins has a sizable local structure factor, when they
 belong to regions with a larger local coordination
 number. In Fig. \ref{f.realspace}(b) we have picked a site
 on a dangling bond attached to a segment of a ladder,
 and calculated its correlations to the rest of
 the system. Compared with Fig. \ref{f.realspace}(a),
 we observe that the correlation
 volume around the chosen spin is considerably
 smaller than the cluster of ``correlated" spins
 identified by the local structure factor.
 This implies that
 spins that are most correlated with the remainder of the
 system are not all correlated with each other.
 The finite-range nature of correlations is also clearly
 shown in the picture, along with the non-monotonic
 decay of correlations with distance. The latter
 is due to the fact that a dangling spin is mostly
 correlated with spins not involved in a local
 singlet state, even over long distances, due
 to the effective long-range couplings discussed
 in Sec. \ref{ladder}. Nonetheless, even in the presence
 of such couplings, correlations decay significantly
 with distance. This picture is to be contrasted
 with the case in which long-range couplings give
 rise to a genuine order-by-disorder phenomenon,
 as for instance in the site-diluted
 dimerized systems treated in Ref. \onlinecite{Yasudaetal01B}.
 There the real-space image of $C(0,i)$
 shows very little spatial decay of the correlations between
 the free moments of the system.

\subsection{Griffiths-McCoy singularities}

In this section,
the nature of the low-lying excitation spectrum
in the quantum-disordered phase is discussed.
In a clean system, the presence
of a finite real-space correlation length
implies finite correlations also in imaginary
time, i.e. the existence of a gap in the
excitation spectrum. In disordered systems, however, this is
not necessarily the case. The disorder considered
here is completely
uncorrelated in real space but perfectly
correlated in imaginary time, such that long-range
correlations in the time dimension (and thus
absence of a gap) may coexist with short-range
real-space correlations.

\begin{figure}[h]
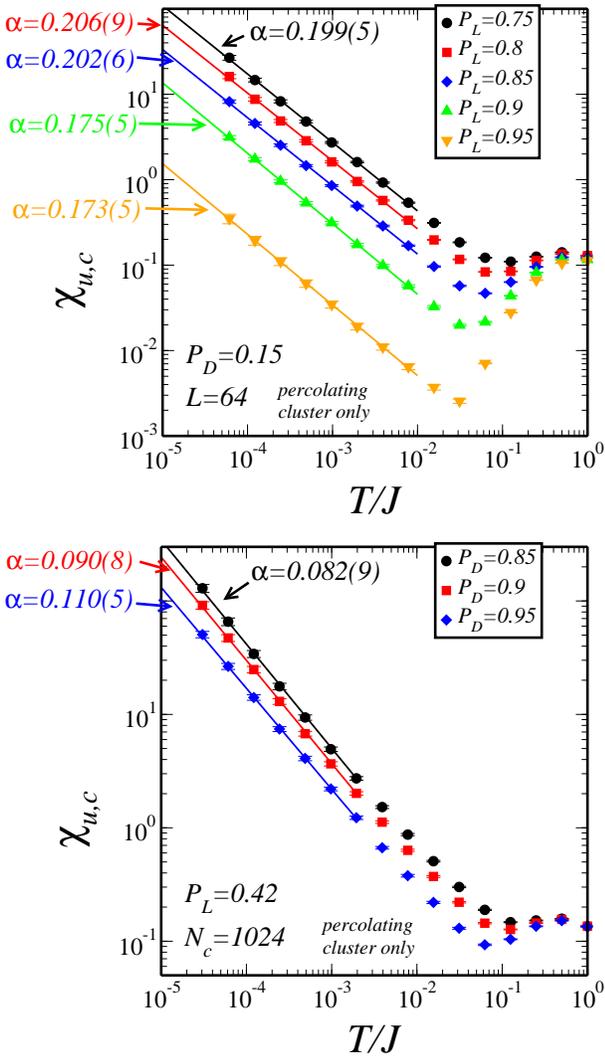

\begin{center}
\includegraphics[
%bbllx=30pt,bblly=20pt,bburx=195pt,bbury=250pt,%
     width=80mm,angle=0]{Fig15A.eps}
\vskip .3cm
\includegraphics[
%bbllx=10pt,bblly=10pt,bburx=250pt,bbury=300pt,%
     width=80mm,angle=0]{Fig15B.eps}
\caption{\label{f.chiu} (color online) Low-temperature uniform
susceptibility of the percolating cluster in the quantum-disordered
regime: ladder regime (upper panel) and dimer regime (lower panel).
The solid lines are power-law fits of the form $\chi_{u,c}\sim
T^{-1+\alpha}$, and the resulting fit coefficients $\alpha$ are
indicated. In the upper panel, the percolating cluster is picked as
the largest cluster in a $64\times 64$ lattice, whereas in the lower
panel it is grown freely from a seed site up to a fixed site of
$N_c= 1024$ (see Sec. \ref{numerics}).}
\end{center}
\end{figure}

 The nature of the low-lying excitation spectrum is
reflected in the low-$T$ uniform
susceptibility. The global uniform susceptibility
of the system is typically dominated by the
Curie contribution from odd-numbered
finite clusters, which masks the
behavior of the quantum-disordered percolating
cluster. Hence, in the following discussion the
contributions from the finite clusters are neglected.
Instead, we focus
on the uniform susceptibility of the percolating
cluster with an \emph{even} number of
sites $N_c$:
\begin{equation}
\chi_{u,c} = \frac{\beta}{N_c} \sum_{ij\in c }
\langle S_i^{z} S_j^{z} \rangle .
\end{equation}

 The results for $\chi_{u,c}$ as a function
of temperature, both in the ladder regime
and in the dimer regime, are shown in
Fig. \ref{f.chiu}. A gapped spectrum
would imply an exponentially vanishing
susceptibility as $T\to 0$. However,
we find a non-vanishing susceptibility
in this limit, and observe that $\chi_{u,c}$
instead displays a \emph{power-law divergence}
\begin{equation}
\chi_{u,c} \sim T^{\alpha-1}
\end{equation}
with $\alpha\neq 0$, such that this divergence
is not Curie-like.  In particular, $\alpha = \alpha({\bm P})$
varies continuously as one scans across the
quantum-disordered region: $\alpha \approx 0.09 \div 0.2$.
This clearly shows
that the system is not only gapless, but that
its gapless excitations lead to singularities
in the response to an external field.

 Quantum-disordered phases with divergent
response properties at $T\to 0$ are familiar in the literature of
disordered systems. They are classified as \emph{quantum Griffiths
phases}, and have been theoretically shown to exist in
bond-disordered dimerized chains \cite{Hymanetal96}, in
bond-disordered $S=1$ (Haldane) chains \cite{HymanY97,Todoetal00B},
and in disordered transverse-field Ising models
\cite{Fisher92,SenthilS96,Pichetal98} (see
Ref.~\onlinecite{IgloiM05} for a recent review). As in our model,
their central feature is the coexistence of short-range correlations
together with so-called Griffiths-McCoy singularities in the
response functions. Such singularities are due to the anomalous role
played by local low-energy excitations living on rare regions of the
system.

\begin{figure}[h]
\begin{center}
\includegraphics[
%bbllx=30pt,bblly=20pt,bburx=195pt,bbury=250pt,%
     width=55mm,angle=0]{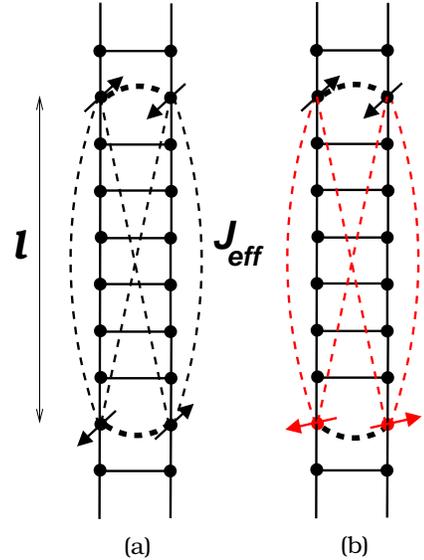}
\caption{\label{f.excite} (color online) Sketch of a low-energy
excitation in the bond-diluted ladder system. (a) Four spins with
missing rung bonds are coupled across a clean region of length $l$
with weak effective couplings $J_{\rm eff} \sim \exp[-\sigma l]$;
(b) rotating the spins at one end of the clean region by the same
angle, spin excitations occur on the weaker bonds only, leaving the
stronger ones unchanged.}
\end{center}
\end{figure}

What are the excitations leading to the singular response in our
model? To answer this question, we focus on the ladder limit,
referring to the discussion of bond dilution in ladder systems given
in the previous section. There, it was observed that bond dilution
leaves free $S=1/2$ moments interacting via effective long-range
couplings, $J_{\rm eff}$, exponentially decaying with the distance.
The presence of this weak energy scale in the system immediately
populates the low-lying spectrum down to the ground state. For any
arbitrarily small energy, one can always find two $S=1/2$ moments
which are sufficiently far from each other to be coupled with this
energy. Nonetheless, to be able to excite the system at this energy
scale only, one needs the two moments to be separated by a large
clean region, such that one can rotate all moments at one end of
this region of a given angle, leaving the higher-energy couplings at
shorter range unchanged (see Fig. \ref{f.excite}). Therefore on the
one hand there is an exponentially small energy of the localized
excitation, leading to an equally weak \emph{local gap} $\Delta$:
\begin{equation}
\Delta \sim \exp[-\sigma l],
\label{e.Delta}
\end{equation}
but, at the same time, the presence of a
large clean region of length $l$ between the free moments
is a \emph{rare event} with exponentially
small probability
\begin{equation}
w(l) \sim (1-P_L)^l \sim \exp[-c~l].
\label{e.prob}
\end{equation}
By substituting Eq. (\ref{e.Delta}) into
Eq. (\ref{e.prob}) with the appropriate
metric term to preserve the normalization
condition of the probability distribution,
the two exponentials compensate each other, leaving
a \emph{power-law} distribution\cite{RiegerY96}
of the local gap $\Delta$:
\begin{equation}
w(\Delta) \sim \Delta^{c/\sigma-1}.
\label{e.probDelta}
\end{equation}

A very similar argument can be applied
to the dimer limit of the model.
Here, \emph{e.g.}, an exponentially rare
long clean decorated chain with
weakly interacting spins at its ends
leads to a similar probability
distribution of local gaps.

 The above result has immediate consequences
for the thermodynamics of the system \cite{Fisher92},
giving rise to the observed power-law behavior of the
uniform susceptibility $\chi \sim T^{\alpha-1}$
where $\alpha = c/\sigma$. In particular
the constant $c$ is clearly disorder-dependent
[$c = |\ln(1-P_L)|$ via  Eq.~(\ref{e.prob})],
such that $\alpha$ is expected to be non-universal,
as it is observed in our data.

\begin{figure}
\begin{center}
\includegraphics[%bbllx=0pt,bblly=20pt,bburx=328pt,bbury=259pt,%
     width=90mm,angle=0]{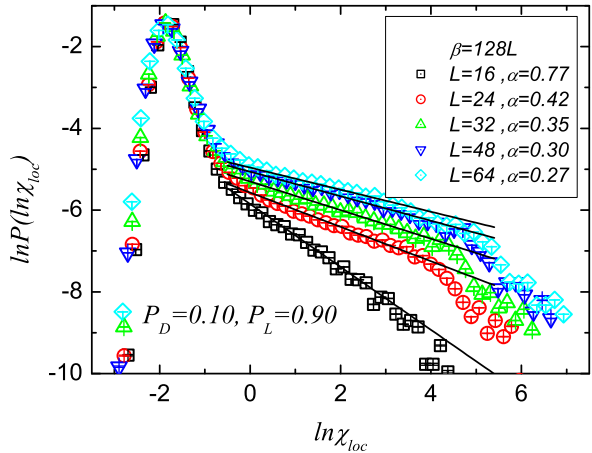}
\caption{\label{f.dist.chi.loc} (color online) Probability
distribution of the local susceptibility for a representative point
in the quantum-disordered phase ($P_D=0.15$, $P_D=0.9$), taken at
$\beta=128 L$. The slope of  the linear tail of the distribution is
proportional to the $\alpha$ exponent in the thermodynamic limit
(see text).} \vskip -.5cm
\end{center}
\end{figure}

 In order to verify the microscopic interpretation
of the QMC results for $\chi_{u,c}$ in terms of a
power-law distribution of the local gaps,
Eq.~(\ref{e.probDelta}), we study
an observable that directly probes the local gap, i.e.
the \emph{local susceptibility} \cite{RiegerY96},
\begin{equation}
\chi_{_{{\rm loc},i}} \equiv \int^{\beta}_0 d\tau \langle S^{z}_i(\tau)S^{z}_i(0)
\rangle. \label{e.chi_loc}
\end{equation}
Hence, if the $i$-th spin is involved in a local
quantum state with a local gap $\Delta$, one
obtains in the large-$\tau$ (low-$T$) limit:
\begin{equation}
\langle S^{z}_i(\tau)S^{z}_i(0) \rangle \sim \exp[-\Delta\tau],
\end{equation}
such that, upon imaginary-time integration
\begin{equation}
\chi_{_{\rm loc},i} \sim \frac{1}{\Delta}.
\end{equation}
By a simple change of variables in Eq.~(\ref{e.probDelta})
one then finds a probability distribution
for the local susceptibility,
\begin{equation}
w(\chi_{_{\rm loc}}) \sim \chi_{_{\rm loc}}^{-1-\alpha},
\end{equation}
which is more conveniently recast in the
form
\begin{equation}
\ln w (\ln \chi_{_{\rm loc}}) \sim -\alpha \ln \chi_{_{\rm loc}},
\label{e.logw}
\end{equation}
i.e. the function $\ln w (\ln \chi_{_{\rm loc}})$ has
a tail linear in $(\ln \chi_{_{\rm loc}})$ and
a slope that directly yields the exponent $\alpha$.

 Fig. \ref{f.dist.chi.loc} shows results for the
logarithmic distribution of the local susceptibility,
Eq.~(\ref{e.logw}), on sites of the percolating cluster only,
for a representative point
in the quantum-disordered phase ($P_L = 0.9, P_D = 0.1$)
and for various system sizes
at very low temperatures ($\beta = 128 L$).
Consistent with the Griffiths-singularity scenario,
one observes a linear tail in $\ln w (\ln \chi_{_{\rm loc}})$,
reflecting the power-law tail of $w(\chi_{_{\rm loc}})$
for all considered system sizes. This tail shows
a strong size dependence, with a decreasing
slope for increasing size, and crossing
over into a non-linear finite-size tail for
large $\chi_{_{\rm loc}}$ values. The tail
of the distribution $w(\chi_{_{\rm loc}})$ is
the result of rare events associated with clean
regions and exponentially small gaps,
such that finite-size corrections
significantly affect the tail statistics.
Nonetheless, the extracted values of the
tail slope $\alpha$ converge towards the
range $\alpha \approx 0.2\div 0.3$,
fully consistent with the independent
estimate obtained from the
low-temperature
behavior of the global susceptibility
$\chi_{u,c}$ in the ladder limit.
This result completes the picture of a
quantum Griffiths phase for the quantum-disordered
region of the system.

\section{\label{conclusion}Conclusions}

In this work, we have investigated
the highly non-trivial interplay between
quantum fluctuations and geometric disorder
realized in the inhomogeneously bond-diluted
quantum Heisenberg antiferromagnet on the
square lattice. We have observed that the
inhomogeneous nature of bond dilution
allows for a continuous tuning of
quantum fluctuations from \emph{linear}
to \emph{non-linear}. Inhomogeneity enables
us to tune the system from a renormalized
classical phase, in which the magnetic
transition is purely driven by geometric
percolation, to a genuine quantum regime
in which the magnetic transition is
decoupled from the percolation transition.
Hence, a novel quantum-disordered phase appears
between the magnetic and
the percolation transition. This phase
has the nature of a quantum Griffiths
phase, characterized by finite-range
correlations coexisting with a divergent
uniform susceptibility.

This represents, to our knowledge, the first evidence of a quantum
Griffiths phase in disordered \emph{2D Heisenberg} antiferromagnets
\cite{footnote}, the previous observations being limited so far to
one-dimensional Heisenberg antiferromagnets
\cite{Hymanetal96,HymanY97,Todoetal00B} and one- and two-dimensional
quantum Ising models
\cite{Fisher92,SenthilS96,Pichetal98,RiegerY96}.
%v2!
 Given the relevance of the
two-dimensional Heisenberg model to the physics of cuprate oxides,
we believe that this finding represents an important step towards
the possibility of the experimental observation of a quantum
Griffiths phase.

 Obviously, the most natural form of disorder
 in real Heisenberg antiferromagnets is site
 dilution, realized by doping of non-magnetic
 impurities in the magnetic lattice \cite{Vajketal02}.
 Bond dilution can, in principle, be realized
 by vacancies in the non-magnetic lattice,
 in particular by vacancies on the sites
 of the non-magnetic ions involved in the
 superexchange paths. The presence of
 such vacancies naturally alters the local
 electronic structure of the system, such
 that its effect on the magnetic Hamiltonian
 may be more elaborate than simple bond
 dilution, and it might even introduce excess
 charge carriers in the system. The doping
 of the non-magnetic ions involved in the
 superexchange paths with other non-magnetic
 species introduces instead bond disorder
 \cite{bonddisorder}, whose effect can in principle
 be arbitrarily close to that of bond dilution.

 Beside the technical challenges in
realizing bond dilution (and bond disorder
in general) in real antiferromagnets,
a fundamental aspect of this work is the
central role of inhomogeneity in the
search for novel quantum phases
induced by disorder in two-dimensional
Heisenberg antiferromagnets. Recent
studies \cite{Laflorencie05} show
that the N\'eel ordered state
of the square-lattice Heisenberg antiferromagnet
is extremely stable towards \emph{homogeneous} bond
disorder, and homogeneous bond dilution destroys
N\'eel order only at the classical percolation
threshold \cite{Sandvik02A}. Hence,
inhomogeneity is an
essential ingredient to realize strong
enhancement of quantum effects through
disorder beyond the previously investigated
scenarios.
\section{\label{section8}Acknowledgments}

We acknowledge fruitful discussions with L. Balents, N. Bray-Ali, N.
Laflorencie, B. Normand, G. Refael, S. Sachdev, H. Saleur and A.
Sandvik. S. H. and R. Y. acknowledge hospitality at the KITP, Santa
Barbara. This work is supported by the DOE award DE-FG02-05ER46240.
R. Y. was also supported in part by the NSF through grant No.
PHY99-07949. Computational facilities have been generously provided
by the HPCC-USC Center.


\begin{thebibliography}{200}
\bibitem{footnote2} A short, partial account of this work has appeared
previously \cite{Yuetal05}.
\end{thebibliography}
\end{document}